\documentclass[a4paper,11pt]{article}
\pdfoutput=1 
\usepackage{jcappub}
\usepackage{amsmath,amssymb}
\usepackage{bm}
\usepackage{verbatim}
\usepackage{xcolor}
\usepackage{graphicx}
\usepackage{enumitem}
\usepackage{geometry}
\usepackage{xspace}
\usepackage{multirow}
\usepackage{pdflscape}
\usepackage{placeins}
\usepackage{epstopdf}
\usepackage{comment}
\makeatletter
\gdef\@fpheader{}
\g@addto@macro\bfseries{\boldmath}
\makeatother

\newcommand{\OmegaGW}{\Omega_{\mathrm{GW}}}
\newcommand{\rhoGW}{\rho_{\mathrm{GW}}}

\let\oldsqrt\sqrt
\def\sqrt{\mathpalette\DHLhksqrt}
\def\DHLhksqrt#1#2{%
\setbox0=\hbox{$#1\oldsqrt{#2\,}$}\dimen0=\ht0
\advance\dimen0-0.2\ht0
\setbox2=\hbox{\vrule height\ht0 depth -\dimen0}%
{\box0\lower0.4pt\box2}}

\newcommand{\dd}{\mathrm{d}}

\newcommand{\sss}[1]{{\scriptscriptstyle{#1}}}
\newcommand{\boldmathsymbol}[1]{{\ensuremath{\boldsymbol{#1}}}}

\newcommand{\uPl}{\mathrm{Pl}}

\newcommand{\usssPl}{\sss{\uPl}}

\newcommand{\calH}{\mathcal{H}}

\newcommand{\Mp}{M_\usssPl}

\newcommand{\beq}{\begin{equation}}
\newcommand{\eeq}{\end{equation}}
\newcommand{\bea}{\begin{equation}\begin{aligned}}
\newcommand{\eea}{\end{aligned}\end{equation}}

\newlength{\wsingfig}
\newlength{\wdblefig}
\newlength{\wquadfig}
\newlength{\wtriplefig}
\newcommand{\Eq}[1]{Eq.~(\ref{#1})}

\newcommand{\Fig}[1]{Fig.~{\ref{#1}}}

\newcommand{\Hc}[1]{\mathcal{H}}

\renewcommand{\Hc}{\mathcal{H}}

\def\doi{http://doi.org}

\date{today}

\title{Primordial black holes and gravitational waves from non-canonical inflation}

\author[a]{Theodoros Papanikolaou}

\author[b]{Andreas Lymperis}

 \author[b]{Smaragda Lola}

\author[a,c,d]{Emmanuel N. Saridakis}

\affiliation[a]{National Observatory of Athens, Lofos Nymfon, 11852 Athens, 
Greece}

\affiliation[b]{Department of Physics, University of Patras, 26500 Patras, Greece}

\affiliation[c]{CAS Key Laboratory for Researches in Galaxies and Cosmology, 
Department of Astronomy, University of Science and Technology of China, Hefei, 
Anhui 230026, P.R. China}
 
\affiliation[d]{Departamento de Matem\'{a}ticas, Universidad Cat\'{o}lica del Norte, Avda.
Angamos 0610, Casilla 1280 Antofagasta, Chile}

\emailAdd{papaniko@noa.gr}
\emailAdd{alymperis@upatras.gr}
\emailAdd{magda.lola@upatras.gr}
\emailAdd{msaridak@noa.gr}

\abstract{Primordial black holes (PBHs) can generically form in inflationary setups through the collapse of enhanced cosmological perturbations, 
providing us
access to the early Universe through their associated observational signatures. 
In the current work we propose a new mechanism of PBH production within non-canonical inflation,
using a class of steep-deformed inflationary potentials compatible with natural values for the non-canonical exponents.
In particular,   requiring   significant PBH production
we extract constraints on the non-canonical exponents. Additionally,
we find that our scenario can lead to the formation of    asteroid-mass 
PBHs,
which can account for the totality of the dark matter, as well as to production of solar-mass PBHs within the LIGO-VIRGO detection band. 
Finally, we find that the  enhanced cosmological perturbations which collapse to form PBHs can produce
a stochastic gravitational-wave (GW) background induced by second-order gravitational interactions. Very interestingly, we obtain a GW signal detectable by future GW experiments, in particular by SKA, LISA and BBO.
}


\begin{document}
\maketitle


\section{Introduction}\label{sec:intro}

Inflation~\cite{Starobinsky:1980te, Guth:1980zm, Linde:1981mu,
Albrecht:1982wi, Linde:1983gd,Kazanas:1980tx} constitutes one of the most promising paradigm since it can describe the physical conditions prevailed in the very early universe and explain fundamental problems of the Big-Bang cosmology, namely the flatness and the horizon problems. During the inflationary era, the Universe underwent a phase of accelerated expansion which, within the simplest inflationary setups, is driven by a scalar field, the inflaton, slowly rolling down its potential. Then, at some point inflation ends with the inflaton field oscillating at the bottom of its potential and decaying into other degrees of freedom it couples to~\cite{Albrecht:1982mp, Dolgov:1982th, Abbott:1982hn, Turner:1983he, Shtanov:1994ce, Kofman:1994rk, Kofman:1997yn} . Finally, after the thermalisation of the decay products, one is met with the onset of the Hot Bing Bang (HBB) radiation-dominated (RD) era.

Traditionally, the majority of inflationary mechanisms was studied through the introduction of scalar fields with canonical kinetic terms.
However, this needs not be the case. In fact, several problems encountered in inflation, including fine-tuning issues due to tiny dimensionless constants, as well as large predictions for tensor fluctuations which can naturally be resolved in theories of scalar fields with non-canonical kinetic terms 
\cite{Armendariz-Picon:1999hyi,
Garriga:1999vw,Mukhanov:2005bu,
Barenboim:2007ii, Tzirakis:2008qy, Franche:2009gk,
Unnikrishnan:2012zu, Gwyn:2012ey, Zhang:2014dja,Cai:2014uka,
Gwyn:2014wna, Hossain:2014xha, Rezazadeh:2014fwa,Sheikhahmadi:2016wyz,Geng:2017mic,
Dimopoulos:2017zvq, Mohammadi:2018wfk,Benisty:2019jqz,
Kamenshchik:2018sig,Karydas:2021wmx,Lymperis:2021emd}. Such terms arise most naturally in supergravity and string compactifications, which typically contain a large number of light scalar fields (moduli), with dynamics that are governed by a non-trivial moduli space metric $G_{ij}$ ~\cite{
Freedman:1976xh, Cremmer:1982en, Nilles:1983ge, 
Green:1987sp}. As long as the moduli space metric is not flat, we generically expect non-canonical kinetic terms, which can have significant cosmological consequences. Among others, additional friction terms in the equations of motion of the inflaton slow down the scalar field, thus significantly reducing the resulting tensor-to-scalar ratio without ruining the spectral index
\cite{Armendariz-Picon:1999hyi,
Garriga:1999vw,Mukhanov:2005bu,
Barenboim:2007ii, Tzirakis:2008qy, Franche:2009gk,
Unnikrishnan:2012zu, Gwyn:2012ey, Zhang:2014dja,Cai:2014uka,
Gwyn:2014wna, Hossain:2014xha, Rezazadeh:2014fwa,Sheikhahmadi:2016wyz,Geng:2017mic,
Dimopoulos:2017zvq, Mohammadi:2018wfk,Benisty:2019jqz,
Kamenshchik:2018sig,Karydas:2021wmx}.

On the other hand, primordial black holes (PBHs), firstly proposed in the early `70s~\cite{1967SvA....10..602Z, Carr:1974nx,1975ApJ...201....1C,1979A&A....80..104N}, constitute a general prediction of inflationary models which present an enhanced curvature power spectrum on small scales compared to the ones probed by Cosmic Microwave Backgroun (CMB) and Large Scale Structure (LSS) experiments [See here~\cite{Khlopov:2008qy,Carr:2020gox} for nice reviews on the topic]. In particular, PBHs have rekindled the interest of the scientific community since, among others, they constitute a viable candidate for dark matter accounting for a part or the totality of its contribution to the energy budget of the Universe \cite{Chapline:1975ojl,Belotsky:2014kca}. They can additionally explain
the large-scale structure formation through Poisson 
fluctuations \cite{Meszaros:1975ef,Afshordi:2003zb}, providing as well the seeds for the supermassive black holes residing in the galactic centres~\cite{1984MNRAS.206..315C, Bean:2002kx} while at the same time they can probe physical phenomena at very high energy scales~\cite{Ketov:2019mfc}. Interestingly, they can also account for the black-hole merging events recently 
detected by the LIGO/VIRGO collaboration~\cite{LIGOScientific:2018mvr}. Other indications in favor of the existence of PBHs
can be found in \cite{Clesse:2017bsw}.

PBHs are also associated with numerous gravitational-wave (GW) signals, which can be potentially detected by current and future GW experiments. Indicatively, one can mention the stochastic GW background associated to black-hole merging events~\cite{Nakamura:1997sm, Ioka:1998nz, 
Eroshenko:2016hmn,Zagorac:2019ekv, Raidal:2017mfl}, like the ones recently detected by LIGO/VIRGO~\cite{LIGOScientific:2018mvr},  as well the second order GWs induced from primordial curvature 
perturbations~\cite{Bugaev:2009zh, Saito_2009, Nakama_2015, 
Yuan:2019udt,Domenech:2021and,Balaji:2022rsy} (for a recent 
review see \cite{Domenech:2021ztg}) or from Poisson PBH energy density 
fluctuations~\cite{Papanikolaou:2020qtd,Domenech:2020ssp,Papanikolaou:2022chm}. Interestingly, one should highlight that through the aforementioned GW portal PBHs can act as well as a novel probe constraining modified gravity theories~\cite{Papanikolaou:2021uhe,Papanikolaou:2022hkg,Kawai:2021edk}. 

Thus, given the above mentioned motivation regarding non-canonical inflation, the rekindled interest on PBHs as well as the huge progress witnessed in the recent years in the field of GW astronomy, there has been witnessed during the last years an increasing interest in the literature bridging the above mentioned fields together~\cite{Kamenshchik:2018sig,Fu:2019ttf,Lin:2020goi,Yi:2020cut,Solbi:2021wbo,Solbi:2021rse,Heydari:2021gea,Heydari:2021qsr,Teimoori:2021pte,Ahmed:2021ucx}. In this paper, based on our previous work~\cite{Lola:2020lvk}, we study as well PBH formation within non-canonical inflation focusing on a class of steep-deformed inflationary potentials~\cite{Geng:2015fla} which is in general in remarkable agreement with the Planck data and compatible as well with natural values for the non-canonical exponents. Interestingly, this class of inflationary potentials can exhibit an inflection point making in this way the inflaton field rolling down an almost flat inflationary plateau and enhancing in this way the curvature power spectrum on small scales compared to the CMB and the LSS ones. As a consequence, one is able to produce PBHs and extract subsequently the GW signals associated to them.

The paper is structured as follows: In Section \ref{sec:non_canonical_inflation} we review the basic equations of a single-field inflationary theory with non-canonical kinetic terms and a steep-deformed potential. Within this framework, we study the dynamical equations of the scalar field and the resulting perturbations. Then, we dedicate Section \ref{sec:PBH} to the study of primordial black hole formation within our model deriving as well the contribution of PBHs to dark matter. Followingly, in Section \ref{sec:GW} we investigate the gravitational waves induced at second order from enhanced curvature perturbations which collapsed to form PBHs. Finally, Section \ref{sec:conclusions} is devoted to conclusions.

\section{Non-Canonical Inflation with a steep-deformed potential}\label{sec:non_canonical_inflation}
We study here an inflationary theory with non-canonical kinetic terms with a well theoretically justified Lagrangian~\cite{Armendariz-Picon:1999hyi,Garriga:1999vw,Mukhanov:2005bu,Li:2012vta,Unnikrishnan:2008ki} described by the following action:
\begin{equation}\label{eq:action}
  S = \frac{1}{16\pi G} \int \mathrm{d}^4x \sqrt{-g} \left[ X \left(\frac{X}{M^4}\right)^{\alpha - 1} - V(\phi)\right],
\end{equation}
where $X=-\frac{1}{2}\partial_\mu\phi\partial^\mu\phi$ is the kinetic energy of the scalar field and $V(\phi)$ is the inflationary potential. The parameter $M$ has dimensions of mass and determines the scale in which the non-canonical effects become significant, while $\Mp$ is the reduced Planck mass. Finally, the parameter $\alpha$ is a dimensionless non-canonical parameter quantifying deviations from canonicality. For $\alpha=1$ as one may see from \Eq{eq:action} one recovers the canonical Lagrangian.

\begin{figure*}[t!]
\begin{center}
\includegraphics[width=0.595\textwidth]{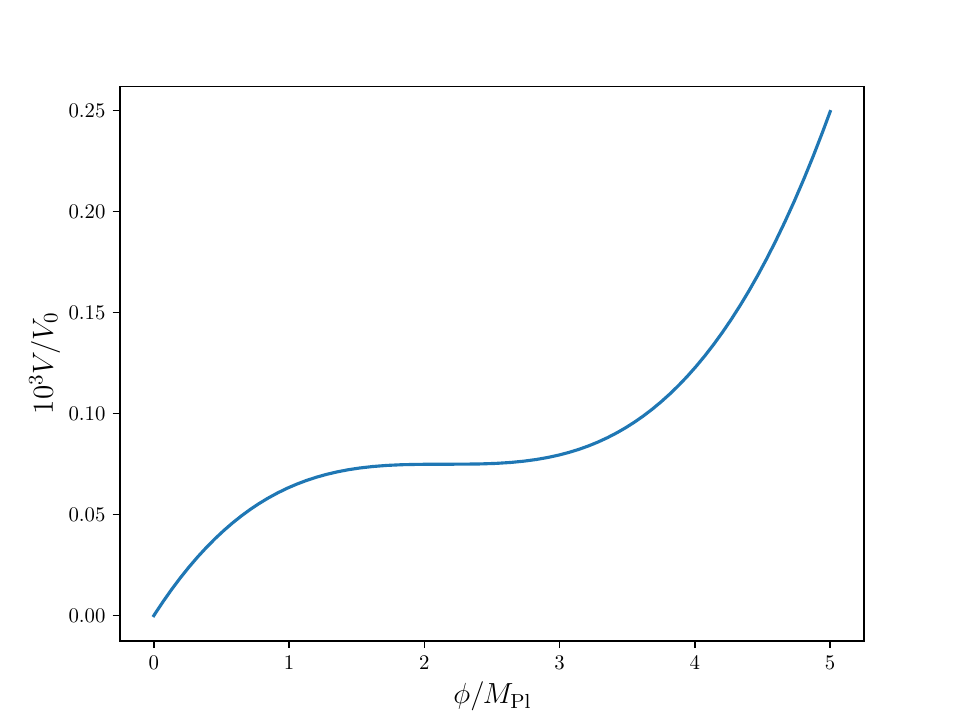}
\caption{{\it{The steep-deformed inflationary 
potential (\ref{eq:V_steepness}), for $n=3$, $V_0=10^{-16}\Mp^4$ and $\lambda = 7.54\times 10^{-6}$ and $\phi_0 = 2.27\Mp$.}}}
\label{fig:V_n_3_a_1_5}
\end{center}
\end{figure*}

In our considerations, we will consider an inflationary potential which has an inflection point, a feature which generically leads to an enhancement of the curvature power spectrum on small scales and PBH formation. In particular, we will focus on a class of steep-deformed inflationary potentials which are in agreement with the Planck data and compatible as well with natural values for the non-canonical exponent $\alpha$~\cite{Lola:2020lvk}. They are also compatible with the standard thermal history of the Universe, with a late-time cosmic acceleration stage in agreement with observations~\cite{Geng:2015fla}. In particular, the class of the inflationary potentials we will start with is of the following form \cite{Geng:2017mic}: 
\beq\label{eq:V_steepness_general}
V(\phi) = V_0e^{-\lambda \phi^n/\Mp^n},
\eeq
with $V_0$ and $\lambda$ being the usual potential parameters and $n$ the exponent parameter that determines the deformed-steepness. As standardly adopted in the literature, we want an inflationary potential which vanishes at $\phi = 0$ in agreement with the fact that the current cosmological constant is negligible in comparison with the energy density stored in the scalar field during inflation. Consequently, making a proper shift transformation the potential form we will work with will be of the following form:
\beq\label{eq:V_steepness}
V(\phi)= V_0e^{-\lambda(\phi_\mathrm{0}-\phi)^n/\Mp^n}- V_0e^{-\lambda\phi^n_\mathrm{0}/\Mp^n},
\eeq
where we have subtracted the last term in order to obtain
 $V(0) = 0$. In order  to have an exact inflection point at $\phi=\phi_0$ one should require that $n\geq 3$. For transparency, in \Fig{fig:V_n_3_a_1_5} we present the inflationary potential for some representative values of the model parameters, namely for $n=3$, $V_0=10^{-16}\Mp^4$, $\lambda = 7.54\times 10^{-6}$ and $\phi_0 = 2.27\Mp$.

\subsection{Background evolution}
In the case of a flat Friedmann-Lema\^itre-Robertson-Walker (FLRW) background, where $\mathrm{d}s^2 = -\mathrm{d}t^2 +a^2(t)\mathrm{d}x^i\mathrm{d}x_i$, the Friedmann equations read as 

\begin{align}\label{eq:Friedmann}
H^2 & = \frac{1}{3\Mp^2}\left[(2\alpha - 1)X\left(\frac{X}{M^4}\right)^{\alpha -1} + V(\phi)\right] \\ 
\dot{H} & = -\frac{1}{\Mp^2}\alpha X\left(\frac{X}{M^4}\right)^{\alpha -1}.
\end{align}
By minimising now the action (\ref{eq:action}) one can straightforwardly obtain the Klein-Gordon (KG) equation for the background evolution of the scalar field $\phi$ which can be recast as follows:
\begin{equation}\label{eq:KG - non-canonical case}
\ddot{\phi}+\frac{3H\dot{\phi}}{2\alpha -1} + \frac{V^\prime(\phi)}{\alpha(2\alpha -1)}\left(\frac{2M^4}{\dot{\phi}^2}\right)^{\alpha - 1} = 0.
\end{equation}
Note that one can write the above equation 
in the form of the energy density conservation
equation $\dot{\rho} + 3H(\rho + p) = 0$ 
with $\rho$ and $p$ being the energy density 
and the pressure of the scalar field $\phi$ which 
read like
\begin{eqnarray}
    \rho & = & (2\alpha-1)X\left(\frac{X}{2M^4}\right)^{\alpha - 1} + V(\phi) \\
     p & = & X\left(\frac{X}{2M^4}\right)^{\alpha - 1} - V(\phi).    
\end{eqnarray}

Using now the definition of the first slow-roll parameter $\epsilon_1\equiv  -\dot{H}/H^2$ and redefine the time variable as the e-fold number defined as $N=\ln \left(\frac{a}{a_\mathrm{ini}}\right)$, with the initial scale factor $a_\mathrm{ini}$ being determined from the CMB pivot scale $k_\mathrm{CMB}$, i.e. $a_\mathrm{ini}=k_\mathrm{CMB}/H_\mathrm{ini}$, one can combine \Eq{eq:Friedmann} and \Eq{eq:KG - non-canonical case} to write the KG equation in the following form:
\beq\label{eq:KG_full_non_canonical}
\phi^{\prime\prime} + \left[\frac{3}{2\alpha -1} - \epsilon_1 \right]\phi^\prime + \frac{V_\phi}{V}\left[3\alpha - (2\alpha -1)\epsilon_1 \right]\frac{\phi^{\prime 2}}{2\epsilon_1}=0.
\eeq
For $\alpha=1$, \Eq{eq:KG_full_non_canonical} acquires its canonical limit form~\cite{Ballesteros:2017fsr}:
\beq\label{eq:KG_full_non_canonical}
\phi^{\prime\prime} + \left[3 - \frac{\phi^{\prime 2}}{2\Mp^2}\right]\left(\phi^\prime + \Mp^2\frac{V_\phi}{V}\right)=0,
\eeq
where we have accounted that for $\alpha=1$, $\epsilon_1=\phi^{\prime 2}/(2\Mp^2)$.

\begin{figure*}[ht]
\begin{center}
\includegraphics[width=0.495\textwidth]{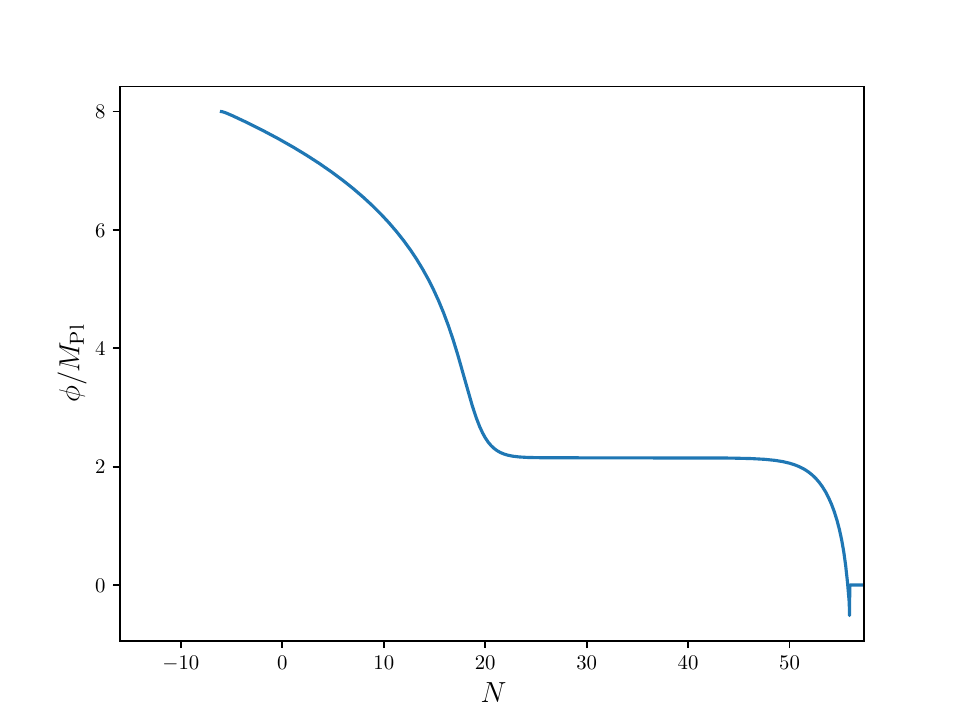}
\includegraphics[width=0.495\textwidth]{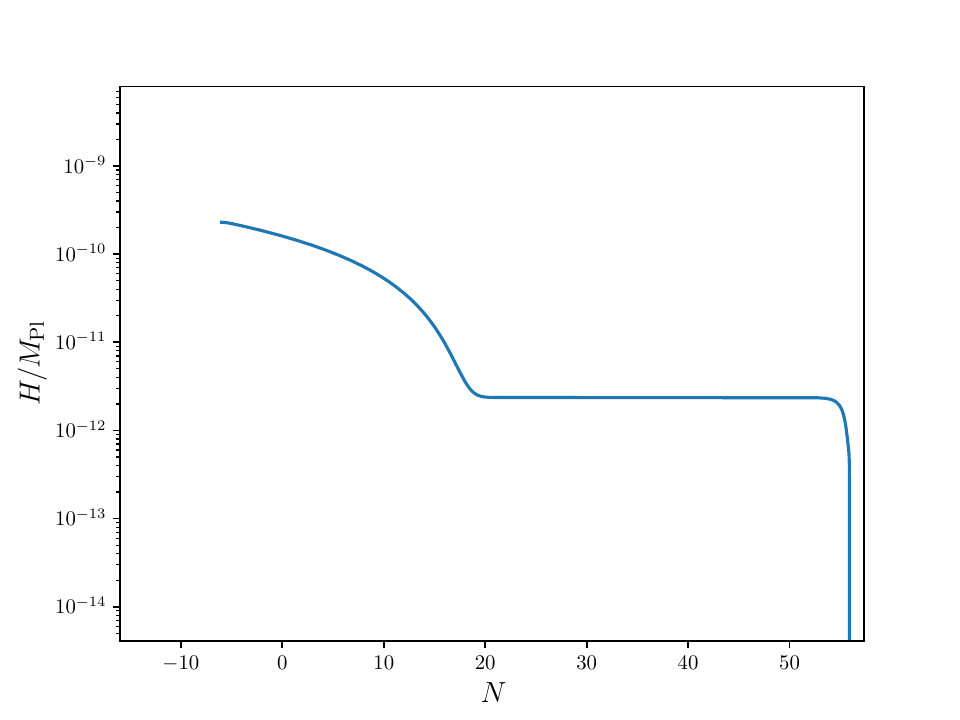}
\caption{{\it{The scalar $\phi$ and the Hubble parameter $H$ as a function of the e-fold number $N$, for $\alpha=1.5$ and $M=10^{-6}\Mp$ and for the inflationary potential of \Fig{fig:V_n_3_a_1_5}.}}}
\label{fig:phi_H}
\end{center}
\end{figure*}

\begin{figure*}[ht]
\begin{center}
\includegraphics[width=0.495\textwidth]{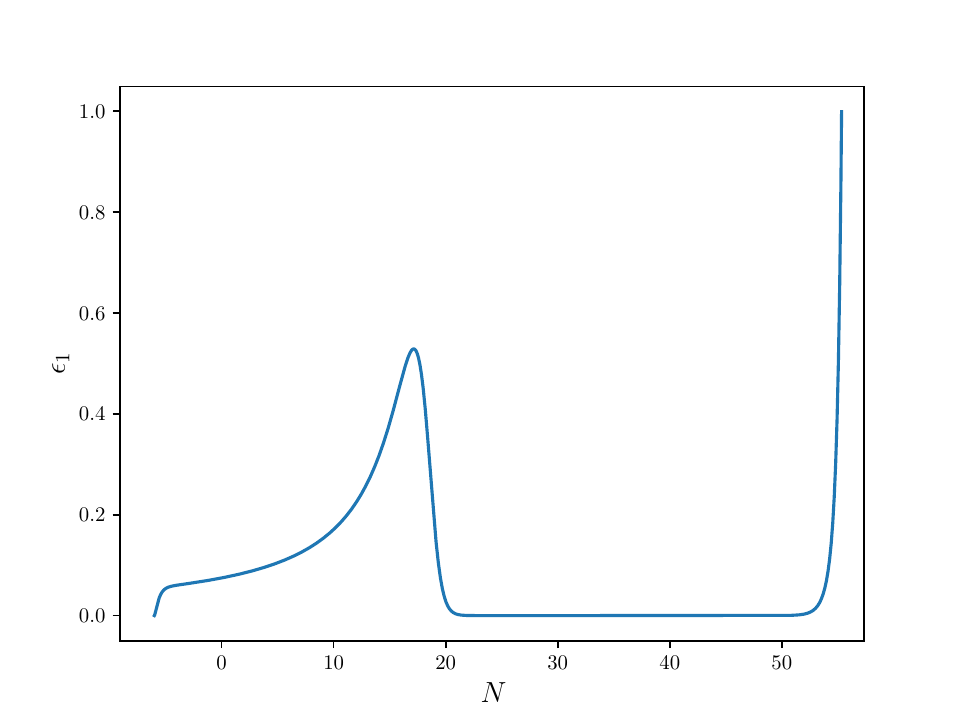}
\includegraphics[width=0.495\textwidth]{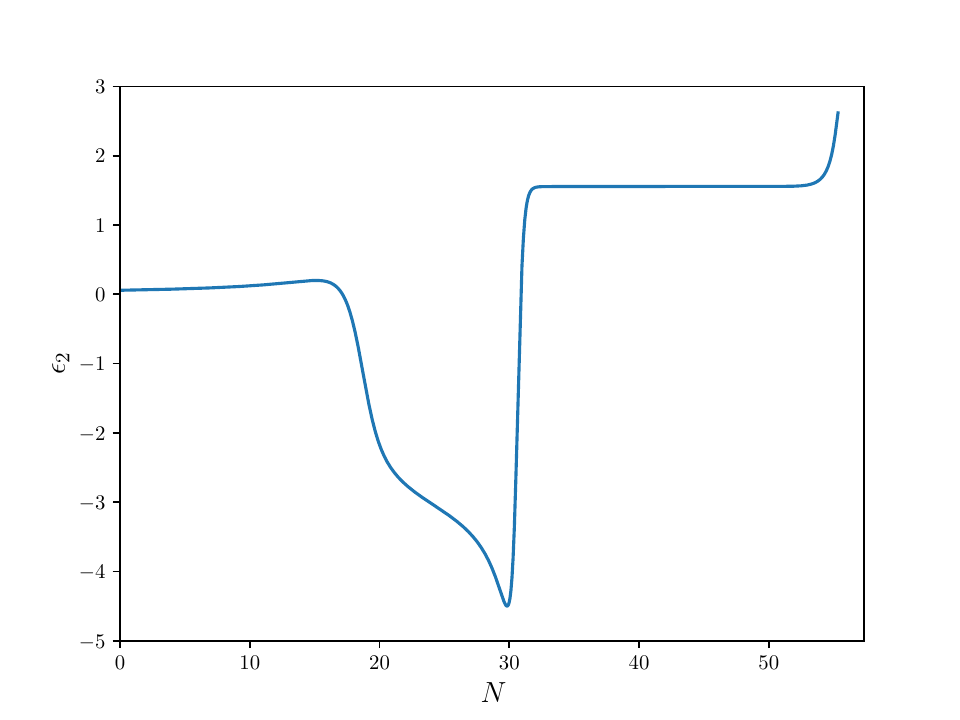}
\caption{{\it{The slow-roll parameters $\epsilon_1$ and $\epsilon_2$ as a function of the e-fold number $N$, for $\alpha=1.5$ and $M=10^{-6}\Mp$ and for the inflationary potential of \Fig{fig:V_n_3_a_1_5}.}}}
\label{fig:epsilon_1_epsilon_2}
\end{center}
\end{figure*}

In \Fig{fig:phi_H} and \Fig{fig:epsilon_1_epsilon_2}
we show the dynamical evolution of the background scalar field $\phi$, of the Hubble parameter as well as of the slow-roll parameters $\epsilon_1$ and $\epsilon_2$ as a function of the e-fold number $N$ for $\alpha=1.5$ and $M=10^{-6}\Mp$. Regarding the parameters of the inflationary potential these are the same as the ones of \Fig{fig:V_n_3_a_1_5}, namely  $n=3$, $V_0=10^{-16}\Mp^4$, $\lambda = 7.54\times 10^{-6}$ and $\phi_\mathrm{0} = 2.27\Mp$ while for the scalar field initial conditions we set $\phi_\mathrm{ini} = 9 \Mp$ and $\phi^\prime_\mathrm{ini} = 8\times 10^{-7}\Mp$.

As one may see, the infaton field is constant at a value around $2\Mp$ for more than $25$ e-folds. This is somehow expected since as we can see from \Fig{fig:V_n_3_a_1_5} the inflationary potential $V(\phi)$ presents an inflection point around this value of the scalar field where $\mathrm{d}V/\mathrm{d}\phi =\mathrm{d}^2V/\mathrm{d}\phi^2 = 0$. In this extremely flat region of the potential, slow-roll (SR) conditions do not hold, and the inflaton field enters into a temporary ultra-slow-roll (USR) period. During this phase, the non-constant mode of the curvature fluctuations, which would decay exponentially in the SR regime, actually grows in the USR regime enhancing in this way the curvature power spectrum at specific scales which can potentially collapse forming PBHs.

\subsection{Perturbations}
\label{sec:scalar}
Having extracting above the background behaviour we study here the perturbations. Focusing on the scalar perturbations, one can write the perturbed FLRW metric in the following form~\cite{Mukhanov:1990me}:
\beq
\begin{split}
\mathrm{d} s^2
= & -(1 + 2\Phi)\mathrm{d}t^2 + 2a(t)\partial_{i} B \mathrm{d}t\mathrm{d}x^i
\\ & + a^{2}(t)\left[(1-2 \Psi)\delta _{ij}+ 2\, \partial_{i}\partial_{j}E\right]\,
\mathrm{d} x^i\, \mathrm{d} x^j,
\end{split}
\eeq
where $\Phi$, $B$, $\Psi$ and $E$ are functions of space and time standing for
the scalar perturbations of the metric. Vector perturbations are neglected here since they are rapidly suppressed during the inflationary stage and therefore they are usually disregarded~\cite{Peter:2013avv}. Finally, the tensor perturbations will be investigated in Section \ref{sec:tensor_perturbations}.

At this point, it is useful to introduce the comoving curvature perturbation $\mathcal{R}$
which is defined as a gauge invariant combination of the metric perturbation $\Psi$
and scalar field perturbation $\delta \phi$, namely
\beq
\mathcal{R} \equiv \Psi + \frac{H}{\dot{\phi}}\delta \phi~.
\eeq
In this way, one can describe in a uniform way both the perturbations of matter and the gravity sector of the Universe. At the end, by 
combining the linearized perturbed Einstein's equation $\delta G^{\mu}_{\;\nu} =  \delta T^{\mu}_{\;\nu}/\Mp^2$ and the equation governing the evolution of the scalar perturbations, it turns out that
\beq\label{eq:R_k}
\ddot{\mathcal{R}}_k +\frac{\dot{z}}{z}\dot{\mathcal{R}}_k+c^2_\mathrm{s}\frac{k^2}{a^2}\mathcal{R}_k = 0,
\eeq
where
\beq
z\equiv \frac{a^3(2Xp_{,X}+4X^2p_{,XX})}{H^2}
\eeq
and $c^{2}_\mathrm{s}$ is the square of the effective speed of sound of the scalar field perturbations defined as~\cite{Garriga:1999vw}
\beq
c^{2}_\mathrm{s} \equiv \left[\frac{\left({\partial {\cal L}}/{\partial X}\right)}{\left({\partial {\cal L}}/{\partial X}\right)
+ 2\, X\, \left({\partial^{2} {\cal L}}/{\partial X^{2}}\right)}\right].
\label{eq:sound_speed_def}
\eeq
For our particular action \eqref{eq:action}, one 
obtains   
$$c^{2}_\mathrm{s} = \frac{1}{2\alpha - 1}.$$
We therefore find that the sound speed is a constant. Since, the square of the sound speed should be a positive number, in the following we  examine the regime $\alpha > 1/2$. We mention that for $0 <\alpha < 1/2$ we acquire superluminal behavior, 
but this is not problematic since it does not  imply pathologies or acausality around a cosmological background \cite{Deffayet:2010qz,Babichev:2007dw}. Nevertheless, for convenience we will focus on the case $\alpha\geq 1$. 
 
At the end, the equation for the comoving curvature perturbations using the e-fold number as the time variable can be recast as
\beq\label{eq:R_k_non_canonical}
\mathcal{R}^{\prime\prime}_k + (3+\epsilon_2-\epsilon_1)\mathcal{R}^\prime_k + \frac{1}{2\alpha-1}\frac{k^2}{a^2H^2}\mathcal{R}_k = 0,
\eeq
where prime denotes derivatives with respect to the e-fold number.
Regarding now the power spectrum of the curvature perturbations the latter is defined as
\beq\label{eq:P_R}
\mathcal{P}_{\mathcal{R}}(k) \equiv \left(\frac{k^{3}}{2\pi^{2}}\right)|\mathcal{R}_{_k}|^{2}.
\eeq

\begin{figure*}[!]
\begin{center}
\includegraphics[width=0.795\textwidth]{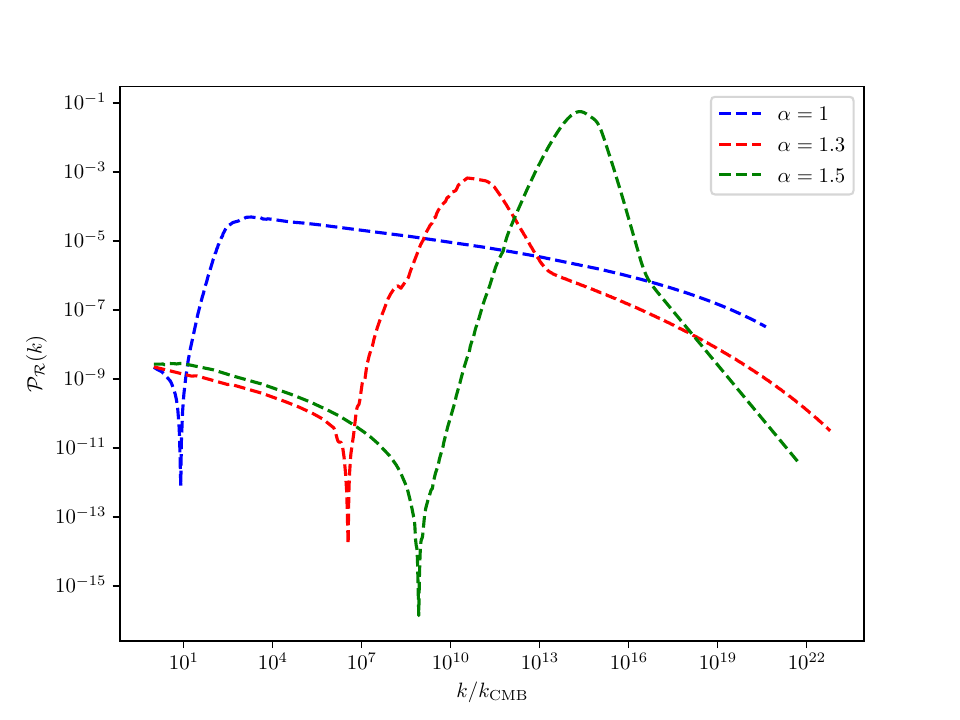}
\caption{{\it{The curvature power spectra for different values of the non-canonical parameter $\alpha$ and for $M=10^{-6}\Mp$. For all curves we fix the inflationary potential parameters to the following values:  $n=3$, $V_0=10^{-16}\Mp^4$, $\lambda = 7.54\times 10^{-6}$ while for the scalar field initial conditions we set $\phi^\prime_\mathrm{ini} = 8\times 10^{-7}\Mp$ and $\phi_\mathrm{ini}=9\Mp$. For the position of the inflection point $\phi_0$ we set $\phi_0 = 1.077\Mp$ (for $\alpha = 1$), $\phi_0 = 1.527\Mp$ (for $\alpha = 1.3$) and $\phi_0 = 2.27\Mp$ (for $\alpha = 1.5$).}}}
\label{fig:P_zeta_vs_alpha}
\end{center}
\end{figure*}
One then can solve numerically \Eq{eq:R_k_non_canonical} using as initial conditions the Bunch-Davis vacuum~\cite{Bunch:1978yq} in the subhorizon regime and plug its solution into \Eq{eq:P_R} in order to extract the curvature power spectrum and see if it can be enhanced at specific scales smaller than the ones probed by the CMB and LSS probes leading in this way to PBH production. In \Fig{fig:P_zeta_vs_alpha} we depict the curvature power spectrum $\mathcal{P}_\mathcal{R}(k)$ for the canonical case $\alpha=1$ as well for the cases where $\alpha =1.3$ and $\alpha = 1.5$. For these three choices of $\alpha$ we fix $\lambda$, $\phi_\mathrm{ini}$ and $n$ and we vary $\phi_\mathrm{0}$ in a way that we maximize the time during which the scalar field stays in the flat region of the potential. In this way, the more time the scalar field stays in the flat region of the potential the more enhancement it will trigger at the level of the curvature power spectrum leading to PBH production.

\begin{figure*}[!]
\begin{center}
\includegraphics[width=0.795\textwidth]{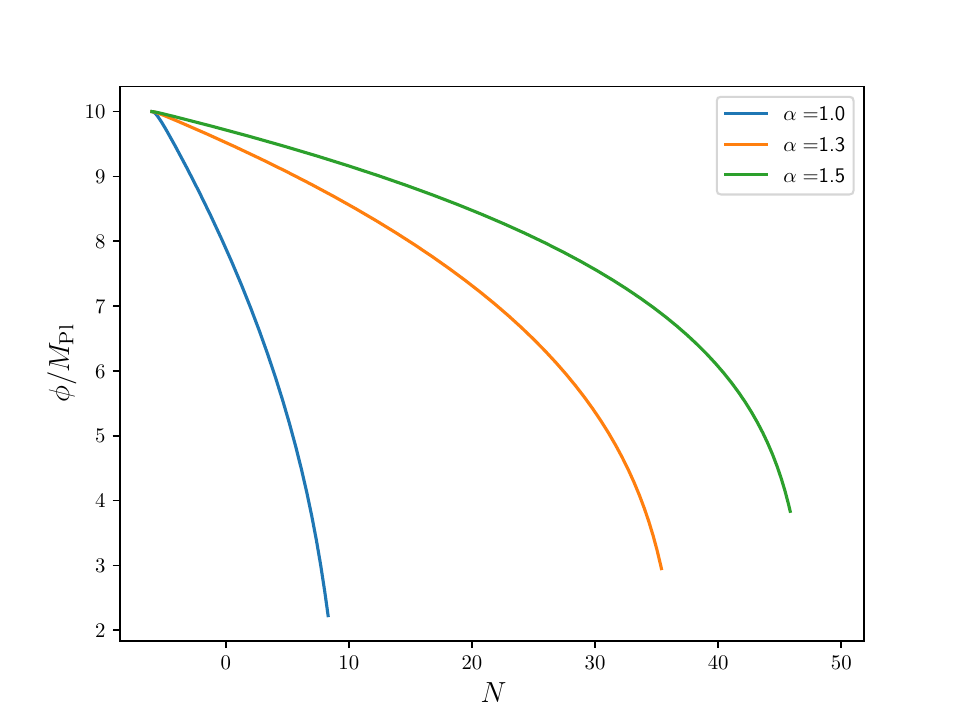}
\caption{{\it{The time evolution of the scalar field $\phi$ for different values of $\alpha$. For all  curves we fix $M=10^{-6}\Mp$ while for the inflationary potential parameters we choose the following values:  $n=3$, $V_0=10^{-16}\Mp^4$, $\lambda = 7.54\times 10^{-6}$, $\phi_0 = \Mp$ with $\phi_\mathrm{ini}=10\Mp$ and $\phi^\prime_\mathrm{ini} = 8\times 10^{-7}\Mp$.}}}
\label{fig:phi_vs_a}
\end{center}
\end{figure*}

In other words, we maximize $\Delta N = N_\mathrm{end}-N_\mathrm{infl}$ where $N_\mathrm{end}$ is the e-fold number at the end of inflation and $N_\mathrm{infl}$ is the e-fold number when the scalar field passes the inflection point and enters the flat region of the inflationary potential. We require at the same time that $\Delta N_* \equiv N_\mathrm{end}-N_\mathrm{CMB}<70$ according to the Planck CMB observational data regarding the number of e-folds elapsed between the time the CMB pivot scale $k_\mathrm{CMB}=0.05\mathrm{Mpc}^{-1}$ exited the Hubble radius during inflation and the end of inflation~\cite{Akrami:2018odb}.

As it can be noticed from \Fig{fig:P_zeta_vs_alpha}, by maximising $\Delta N$ and keeping $\Delta N_*< 70$ one observes an increase of the peak of $\mathcal{P}_\mathcal{R}(k)$ with $\alpha$. This behaviour can be understood with the following reasoning: As it was checked numerically, as $\alpha$ increases the scalar field rolls down its potential slower and slower [See \Fig{fig:phi_vs_a}]. Thus when it reaches its inflection point it is trapped in the flat region of the potential more time compared to the case where the rolling down phase is more abrupt and thus the field acquires a larger velocity so that it can pass the plateau-type region of the inflationary potential. As a consequence, the enhancement of $\mathcal{P}_\mathcal{R}(k)$ is larger for larger values of $\alpha$.

For values of $\alpha$ more or less greater than $1.5$, we find that in order to find a maximum enhancement of the curvature power spectrum at values of the order of $\mathcal{O}(0.1)$ one should require that $\Delta N_*>70$ which is not compatible with the Planck data. This behavior can be understood if we see again \Fig{fig:phi_vs_a} from which we can infer that $\Delta N_*=N_\mathrm{end}$ increases with the increase of $\alpha$. Note that in our setup $N_\mathrm{CMB}=N_\mathrm{ini}=0$.

Varying also the values of $n$ to values larger than $3$ we find that the inflationary potential gets steeper and steeper. Thus, the scalar field acquires a larger velocity $\dot{\phi}$ being able to pass the inflection point and roll down easily up to the end of inflation. In this regime, the scalar field does not stay for a long time within the flat region of the potential suppressing in this way the production of PBHs. Thus, in the following, we will fix the value of $n=3$.

At this point, we should also point out that for $\alpha<1$, the position of the enhancement of the power spectrum takes place around the region of the CMB scales enhancing $\mathcal{P}_\mathcal{R}(k)$ above the value measured by Planck. Thus, regimes where $\alpha<1$ are not considered in what it follows.


\section{Primordial black hole formation}\label{sec:PBH}

We will now study PBH production due to the gravitational collapse of enhanced energy density perturbations that enter the cosmological horizon during the radiation-dominated (RD) era after inflation. In particular, we compute the PBH abundance within the context of peak theory and we compare it with the dark-matter abundance following the general formalism developed in \cite{Young:2019yug}.

Under the assumption of spherical symmetry on superhorizon scales, the overdensity region collapsing to form a black hole is described by the metric~\cite{Starobinsky:1982ee}
\beq\label{eq:metric_spherical}
\mathrm{d}s^2 = -\mathrm{d}t^2 + a^2(t)e^{2\mathcal{R}(r)}\left[\mathrm{d}r^2 + r^2\mathrm{d}\Omega^2\right],
\eeq
where $a(t)$ is the scale factor and $\mathcal{R}(r)$ is the comoving curvature perturbation that is conserved on superhorizon scales~\cite{Wands:2000dp}. Here it is important to note that (\ref{eq:metric_spherical}), which gives the space-time metric after inflation in the non-linear regime when the curvature perturbation $\mathcal{R}$ is not assumed to be small, was first introduced in \cite{Starobinsky:1982ee} without assuming spherical symmetry, and in the form directly relating $\mathcal{R}$ to the difference in the duration of inflation in different points of space in terms of e-folds.  Now, $\mathcal{R}(r)$  is directly linked to the standard energy density contrast in the comoving gauge through the following relation:
\begin{equation}\label{eq:zeta_vs_delta:non_linear}
\begin{split}
\frac{\delta\rho}{\rho_\mathrm{b}} &\equiv \frac{\rho(r,t)-\rho_{\mathrm{b}}(t)}{\rho_{\mathrm{b}}(t)} \\ & = -\left(\frac{1}{aH}\right)^2\frac{4(1+w)}{5+3w}e^{-5\mathcal{R}(r)(r)/2}\nabla^2e^{\mathcal{R}(r)/2},
\end{split}
\end{equation}
where $H(t) = \dot{a}(t)/a(t)$ is the Hubble parameter and $w$ is the equation-of-state parameter $w\equiv p/\rho$.  In the linear regime ($\mathcal{R}\ll 1$) this equation is simplified to
\beq\label{eq:zeta_vs_delta:linear}
\frac{\delta\rho}{\rho_\mathrm{b}}\simeq -\frac{1}{a^2H^2}\frac{2(1+w)}{5+3w}\nabla^2\mathcal{R}(r) \Longrightarrow \delta_k =  -\frac{k^2}{a^2H^2}\frac{2(1+w)}{5+3w}\mathcal{R}_k.
\eeq
Here, the large scales that cannot be observed are naturally removed due to the $k^2$ damping (unlike in $\mathcal{R}$ where the number of PBHs is significantly overestimated, since unobservable scales are not removed when smoothing the PBH distribution~\cite{Young:2014ana}). 

At this point it is important to stress that PBH formation is a non-linear gravitational collapse process. One then should account for the full non-linear relation \eqref{eq:zeta_vs_delta:non_linear} between $\mathcal{R}$ and $\delta$. At the end, one can obtain that the smoothed energy density contrast $\delta_\mathrm{m}$ is related to the linear energy 
density contrast $\delta_l$ defined through \Eq{eq:zeta_vs_delta:linear} by
~\cite{DeLuca:2019qsy,Young:2019yug}
\beq\label{eq:delta_m_smoothed}
\delta_\mathrm{m} = \delta_l - \frac{3}{8}\delta^2_l,
\eeq
where in order to avoid PBH formation on small scales energy density fluctuations are smoothed for scales smaller than the horizon scale  (while larger scales are naturally removed as quoted above). The smoothed linear energy density contrast takes the form
\beq
\delta^R_l = \int \mathrm{d}^3\vec{x}^\prime W(\vec{x},R)\delta(\vec{x}-\vec{x}^\prime) = \int_0^\infty4\pi r^2 W(r,R)\delta(r) \mathrm{d}r,
\eeq
in cartesian and spherical coordinates respectively. The window function $W(\vec{x},R)$ is chosen to be a Gaussian window function whose Fourier transform reads as~\cite{Young:2014ana}
\beq\label{eq:Gaussian_window_function}
\tilde{W}(R,k) = e^{-k^2R^2/2},
\eeq
with the smoothing scale $R$ being equal to the comoving horizon scale $R=(aH)^{-1}$. Using then \Eq{eq:zeta_vs_delta:linear}, the smoothed variance of the energy density field is given by
\beq\label{eq:sigma}
\begin{split}
\sigma^2 & \equiv \langle \left(\delta^{R}_l\right)^2\rangle = \int_0^\infty\frac{\mathrm{d}k}{k}\mathcal{P}_{\delta_l}(k,R)  \\ & = \frac{4(1+w)^2}{(5+3w)^2}\int_0^\infty\frac{\mathrm{d}k}{k}(kR)^4 \tilde{W}^2(k,R) \mathcal{P}_\mathcal{R}(k),
\end{split}
\eeq
where $\mathcal{P}_{\delta_l}(k,R)$ and $\mathcal{P}_{\mathcal{R}}(k)$ denote the reduced energy density and curvature power spectra respectively. 

Regarding the PBH mass, it is of the order of the horizon mass at the horizon crossing time and its spectrum follows a critical collapse scaling law~\cite{Niemeyer:1997mt,Niemeyer:1999ak,Musco:2008hv,Musco:2012au},  
\beq\label{eq:PBH_mass_scaling_law}
M_\mathrm{PBH} = M_\mathrm{H}\mathcal{K}(\delta-\delta_\mathrm{c})^\gamma,
\eeq
where $M_\mathrm{H}$ is the mass within the cosmological horizon at horizon crossing time, and $\gamma \simeq 0.36$ is the critical exponent at the time of PBH formation, here in radiation era.  The parameter $\mathcal{K}$ depends on the equation-of-state parameter and on the shape of the collapsing overdensity. We work with a representative value $\mathcal{K}\simeq 4$~\cite{Musco:2008hv}. 
Regarding the value of the critical threshold $\delta_\mathrm{c}$ this will depend of the shape of the collapsing curvature power spectrum. In our case, as it can be seen from \Fig{fig:P_zeta_vs_alpha} we have broad curvature power spectra. Therefore, in order to compute $\delta_\mathrm{c}$ we will use the treatment of~\cite{Musco:2020jjb} \footnote{Here it is important to stress that in order to compute $\delta_\mathrm{c}$ one should know over which range around the peak of the curvature power spectrum should study the gravitational collapse. To answer this question we need the full non-linear transfer function which demands high-cost numerical simulations going beyond the scope of the present work. Thus, in our analysis we will simply restrict ourselves to modes within the window $k\in [k_\mathrm{peak}/50,50k_\mathrm{peak}]$, where $k_\mathrm{peak}$ corresponds to the position of the peak of $\mathcal{P}_\mathcal{R}(k)$.}.

The PBH mass function $\beta(M)$ can now be evaluated in the context of peak theory. The density of sufficiently rare and large peaks for a random Gaussian density field in spherical symmetry is given by~\cite{Bardeen:1985tr} 
\beq\label{eq:peak_density}
\mathcal{N}(\nu) = \frac{\mu^3}{4\pi^2}\frac{\nu^3}{\sigma^3}e^{-\nu^2/2},
\eeq
where $\nu \equiv \delta/\sigma$ and $\sigma$ is given by \Eq{eq:sigma}. The parameter $\mu$ is the first moment of the smoothed power spectrum defined as
\beq
\begin{split}
\mu^2 & =\int_0^\infty\frac{\mathrm{d}k}{k}\mathcal{P}_{\delta_l}(k,R)\left(\frac{k}{aH}\right)^2 \\ & = \frac{4(1+w)^2}{(5+3w)^2}\int_0^\infty\frac{\mathrm{d}k}{k}(kR)^4 \tilde{W}^2(k,R)\mathcal{P}_\mathcal{R}(k)\left(\frac{k}{aH}\right)^2.
\end{split}
\eeq

The fraction of the energy of the Universe at a peak of a given height $\nu$ which collapses to form a PBH, $\beta_\nu$ is given by 
\beq
\beta_\nu = \frac{M_\mathrm{PBH}(\nu)}{M_\mathrm{H}}\mathcal{N}(\nu)\Theta(\nu - \nu_\mathrm{c})
\eeq
and the total energy fraction of the Universe contained in PBHs of mass $M$ is 
\beq\label{eq:beta_full_non_linear}
\beta(M) = \int_{\nu_\mathrm{c}}^{\frac{4}{3\sigma}}\mathrm{d}\nu\frac{\mathcal{K}}{4\pi^2}\left(\nu\sigma - \frac{3}{8}\nu^2\sigma^2 - \delta_{\mathrm{c}}\right)^\gamma \frac{\mu^3\nu^3}{\sigma^3}e^{-\nu^2/2},
\eeq
where $\nu_\mathrm{c} = \delta_{\mathrm{c},l}/\sigma$ and $\delta_{\mathrm{c},l}=\frac{4}{3}\left(1 -
\sqrt{\frac{2-3\delta_\mathrm{c}}{2}}\right)$.

Having computed before the mass function one can now derive the PBH abundance and its contribution to the dark matter abundance. To do so, we introduce the quantity $f_\mathrm{PBH}$ defined as
\beq\label{eq:f_PBH_def}
f_\mathrm{PBH}\equiv \frac{\Omega_\mathrm{PBH,0}}{\Omega_\mathrm{DM,0}},
\eeq
where the index $0$ refers to today and $\Omega_\mathrm{PBH}=\rho_\mathrm{PBH}/\rho_\mathrm{crit}$ and $\Omega_\mathrm{DM,0} = 0.265$. Accounting now for the fact that PBHs behave like matter one has that 
$\rho_\mathrm{PBH,0}=\rho_\mathrm{PBH,f}\left(a_\mathrm{f}/a_\mathrm{0}\right)^3\simeq \beta \rho_\mathrm{rad,f}\left(a_\mathrm{f}/a_\mathrm{0}\right)^3$ where the index $\mathrm{f}$ refers to PBH formation time and $\beta$ is the PBH mass function [See \Eq{eq:beta_full_non_linear}]. Then, taking into account the fact that the PBH mass is of the order of the mass within the cosmological horizon at PBH formation and applying as well entropy conservation from PBH formation time up to today one gets that  
\beq\label{eq:M_PBH_vs_k}
M_\mathrm{PBH} = M_\mathrm{H}\Omega^{1/2}_\mathrm{rad,0}\left(\frac{g_{*,0}}{g_\mathrm{*,f}}\right)^{1/6}\left(\frac{k_0}{k_\mathrm{f}}\right)^2,
\eeq
where $\Omega_\mathrm{rad}=\rho_\mathrm{rad}/\rho_\mathrm{crit}$ and $g_{*}$ is the number of effective relativistic degrees of freedom. Finally, since $k_\mathrm{f}=a_\mathrm{f}H_\mathrm{f}$ one obtains straightforwardly that 
\beq\label{eq:f_PBH}
f_\mathrm{PBH} = \left(\frac{\beta(M)}{3.27 \times 10^{-8}}\right) \left(\frac{106.75}{g_{*,\mathrm{f}}}\right)^{1/4}\left(\frac{M}{M_\odot}\right)^{-1/2},
\eeq
where $M_\odot$ is the solar mass and where we used the fact that $g_{*,0} = 3.36$~\cite{Kolb:1990vq} and that $\Omega_\mathrm{rad,0}\simeq 10^{-5}$~\cite{Aghanim:2018eyx}. For our numerical applications, we will use $g_{*,\mathrm{f}}=106.75$ since it is the number of effective relativistic degrees of freedom of the Standard Model before the electroweak phase transition~\cite{Kolb:1990vq}.

\begin{figure*}[t!]
\begin{center}
\includegraphics[width=0.496\textwidth]{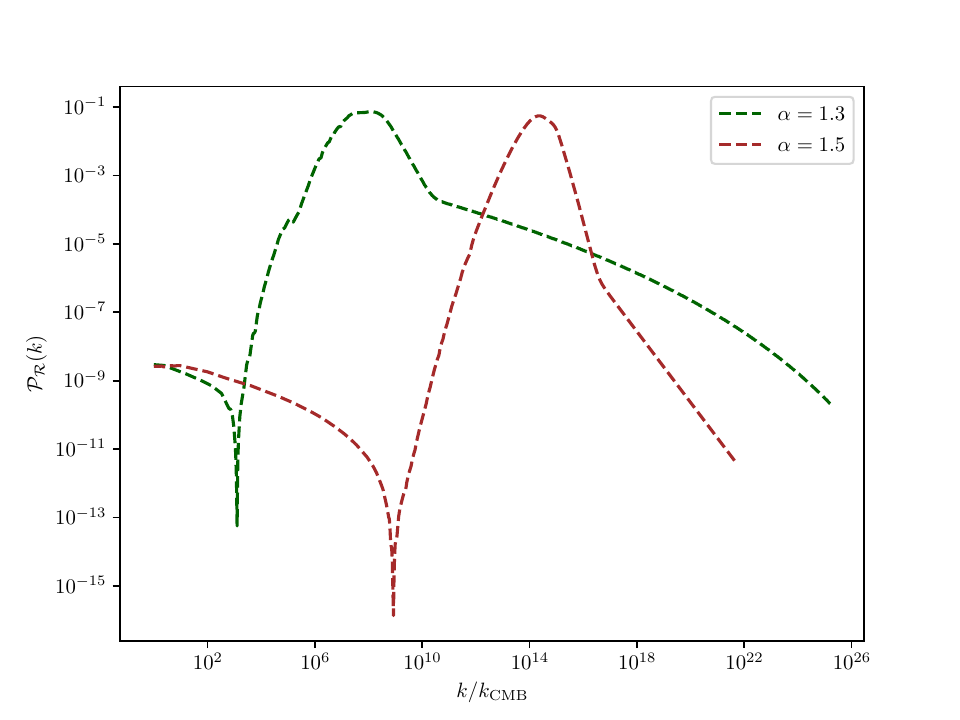}
\includegraphics[width=0.496\textwidth]{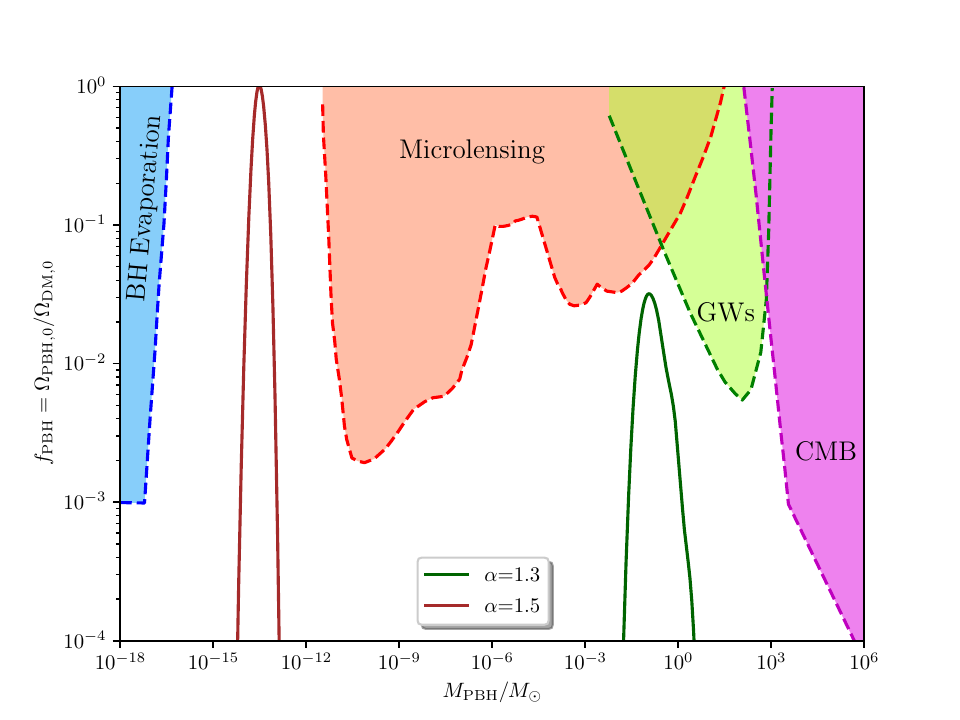}
\caption{{\it{Left Panel: The curvature power spectra $\mathcal{P}_\mathcal{R}(k)$ for $\alpha = 1.3$ and $\alpha = 1.5$ and for the set of   model parameters presented in Table 1. Right Panel: The fraction of dark matter in terms of PBHs denoted as $f_\mathrm{PBH}=\Omega_\mathrm{PBH,0}/\Omega_\mathrm{DM,0}$ as a function of the PBH mass. The colored regions are excluded from evaporation (blue region), microlensing (red region), gravitational-wave (green region) and CMB (violet region) observational probes concerning the PBH abundances. The data for the constraints on $f_\mathrm{PBH}$ from the different observational probes were obtained from~\cite{Green:2020jor}.}}}
\label{fig:f_PBH}
\end{center}
\end{figure*}

In   the left panel  of  \Fig{fig:f_PBH} we depict the curvature power spectra for $\alpha = 1.3$ and $\alpha = 1.5$ and for values of the parameters $\phi_\mathrm{ini}$, $\phi_\mathrm{0}$, $\lambda$ [See Table 1] which can give a power spectrum at the order of $\mathcal{O}(1)$ leading to PBH production. In the right panel on the other hand, we depict the respective fraction of dark matter in terms of PBHs $f_\mathrm{PBH}$ as a function of the PBH mass for the parameters seen in Table 1. In order to derive the PBH mass function (\Eq{eq:beta_full_non_linear}) we computed as well the relevant PBH formation threshold $\delta_\mathrm{c}$ taking into account the broadness of the curvature power spectrum as discussed in~\cite{Musco:2020jjb}[See the 5th row of Table 1]. In the right panel of \Fig{fig:f_PBH} we superimpose as well constraints on $f_\mathrm{PBH}$ from evaporation (blue region), microlensing (red region), gravitational-wave (green region) and CMB (violet region) observational probes constraining the PBH abundances~\cite{Green:2020jor}. For more stringent constraints from BH evaporation see~\cite{Laha:2019ssq,Laha:2020ivk,Saha:2021pqf,Dasgupta:2019cae}.

Interestingly, as we can notice from \Fig{fig:f_PBH} for $\alpha=1.5$ the PBH abundance peaks in the asteroid-mass window $[10^{-17}M_\odot, 10^{-12}M_\odot]$, where PBHs can account for the totality of dark matter. For $\alpha=1.3$ on the other hand, the PBH abundance peaks in the region around one solar mass, which is the order of masses probed by the LIGO-VIRGO GW detectors.

\begin{center}\label{Table 1}
\begin{tabular}{ |p{2cm}||p{2cm}|p{2cm}|  }
\hline
\multicolumn{3}{|c|}{Table 1} \\
 \hline
 & $\alpha = 1.3$  & $\alpha = 1.5$  \\ 
 \hline
 $\phi_0$  & $1.867\Mp$  & $2.27\Mp$ \\ 
 \hline
 $\lambda$  & $5\times 10^{-7}$  & $7.54\times 10^{-6}$ \\ 
 \hline
 $\phi_\mathrm{ini}$  & $10\Mp$  & $9\Mp$ \\ 
  \hline
$\delta_\mathrm{c}$  & $0.443$  & $0.454$ \\ 
  \hline
\end{tabular}
\end{center}
{\vspace *{0.3 cm}}

\section{Gravitational waves}\label{sec:GW}

In the previous section we  studied the production of PBHs within non-canonical inflation. Hence, in this section we can proceed 
to the investigation of the induced gravitational waves generated at second order from the enhanced curvature perturbations which collapsed to form PBHs. 

\subsection{Tensor perturbations}\label{sec:tensor_perturbations}
As a starting point, we consider the Newtonian gauge frame where $\Phi=\Psi$ and $E=B=0$~\footnote{As noted in~\cite{DeLuca:2019ufz, Yuan:2019fwv, Inomata:2019yww,PhysRevLett.113.061301}, the gauge dependence of the tensor modes is expected to disappear in the case of scalar induced gravitational waves generated during a RD era, as the one we study here, due to diffusion damping which exponentially suppresses the scalar perturbations in the late-time limit.}, and by introducing the second-order tensor perturbation $h_{ij}$, to the linearly-perturbed Friedmann-Lema\^itre-Robertson-Walker metric ~\cite{Ananda:2006af,Baumann:2007zm,Kohri:2018awv,Espinosa:2018eve,CANTATA:2021ktz}, the perturbed metric can be written as
\bea
\label{metric decomposition with tensor perturbations}
\mathrm{d}s^2 = a^2(\eta)\left\lbrace-(1+2\Phi)\mathrm{d}\eta^2  + \left[(1-2\Phi)\delta_{ij} + \frac{h_{ij}}{2}\right]\mathrm{d}x^i\mathrm{d}x^j\right\rbrace \, .
\eea
In the Fourier space, the tensor perturbation $h_{ij}$, can be written in terms of the polarisation modes as
\beq
\label{h_ij Fourier decomposition}
h_{ij}(\eta,\boldmathsymbol{x}) = \int \frac{\mathrm{d}^3\boldmathsymbol{k}}{\left(2\pi\right)^{3/2}} \left[h^{(+)}_\boldmathsymbol{k}(\eta)e^{(+)}_{ij}(\boldmathsymbol{k}) + h^{(\times)}_\boldmathsymbol{k}(\eta)e^{(\times)}_{ij}(\boldmathsymbol{k}) \right]e^{i\boldmathsymbol{k}\cdot\boldmathsymbol{x}},
\eeq
where the polarisation tensors $e^{(+)}_{ij}$ and $e^{(-)}_{ij}$ are 
defined as
\begin{eqnarray}
e^{(+)}_{ij}(\boldmathsymbol{k}) \equiv \frac{1}{\sqrt{2}}\left[e_i(\boldmathsymbol{k})e_j(\boldmathsymbol{k}) - \bar{e}_i(\boldmathsymbol{k})\bar{e}_j(\boldmathsymbol{k})\right], \\ 
e^{(\times)}_{ij}(\boldmathsymbol{k}) \equiv \frac{1}{\sqrt{2}}\left[e_i(\boldmathsymbol{k})\bar{e}_j(\boldmathsymbol{k}) + \bar{e}_i(\boldmathsymbol{k})e_j(\boldmathsymbol{k})\right],
\end{eqnarray}
with $e_i(\boldmathsymbol{k})$ and $\bar{e}_i(\boldmathsymbol{k})$ being two three-dimensional vectors which together with $\boldmathsymbol{k}/k$ form an orthonormal basis. At the end, the equation of motion for the tensor modes $h_\boldmathsymbol{k}$ reads as~\cite{Ananda:2006af,Baumann:2007zm,Kohri:2018awv}
\beq
\label{Tensor Eq. of Motion}
h_\boldmathsymbol{k}^{s,\prime\prime} + 2\mathcal{H}h_\boldmathsymbol{k}^{s,\prime} + k^{2}h^s_\boldmathsymbol{k} = 4 S^s_\boldmathsymbol{k}\, ,
\eeq
where 
$s = (+), (\times)$ and the source function $S^s_\boldmathsymbol{k}$ can be recast in the following form:
\beq
\label{Source}
S^s_\boldmathsymbol{k}  =
\int\frac{\mathrm{d}^3 q}{(2\pi)^{3/2}}e^{s}(\boldmathsymbol{k},\boldmathsymbol{q})F(\boldmathsymbol{q},|\boldmathsymbol{k-q}|,\eta)\phi_\boldmathsymbol{q}\phi_\boldmathsymbol{k-q}.
\eeq
In the above expression, we have written the Fourier component of the first order scalar perturbation $\Phi$, usually called as the Bardeen potential, as $\Phi_k(\eta) = T_\Phi(x)\phi_k$ with $x=k\eta$, where $\phi_k$ is the value of the Bardeen potential at some reference initial time $x_0$, which here we consider it to be the horizon crossing time, and $T_\Phi(x)$ is a transfer function, defined as the ratio of the dominant mode of $\Phi$ between the times $x$ and $x_0$. The function $F(\boldmathsymbol{q},|\boldmathsymbol{k-q}|,\eta)$ can be recast in terms of the transfer function as
\bea
\label{F}
F(\boldmathsymbol{q},|\boldmathsymbol{k-q}|,\eta) & = 2T_\Phi(q\eta)T_\Phi\left(|\boldmathsymbol{k}-\boldmathsymbol{q}|\eta\right) 
\\  & \kern-2em + \frac{4}{3(1+w)}\left[\mathcal{H}^{-1}qT_\Phi^{\prime}(q\eta)+T_\Phi(q\eta)\right]\\  & \kern-2em \times \left[\mathcal{H}^{-1}\vert\boldmathsymbol{k}-\boldmathsymbol{q}\vert T_\Phi^{\prime}\left(|\boldmathsymbol{k}-\boldmathsymbol{q}|\eta\right)+T_\Phi\left(|\boldmathsymbol{k}-\boldmathsymbol{q}|\eta\right)\right].
\eea
At the end, \Eq{Tensor Eq. of Motion} can be solved with the Green's function formalism with $h_\boldmathsymbol{k}^{s}$ being given by
\beq
\label{tensor mode function}
h^s_\boldmathsymbol{k} (\eta)  =\frac{4}{a(\eta)} \int^{\eta}_{\eta_\mathrm{d}}\mathrm{d}\bar{\eta}\,  G^s_\boldmathsymbol{k}(\eta,\bar{\eta})a(\bar{\eta})S^s_\boldmathsymbol{k}(\bar{\eta}),
\eeq
where the Green's function  $G^s_{\bm{k}}(\eta,\bar{\eta})$ is the solution of the homogeneous equation 
\beq
\label{Green function equation}
G_\boldmathsymbol{k}^{s,\prime\prime}(\eta,\bar{\eta})  + \left( k^{2} -\frac{a^{\prime\prime}}{a}\right)G^s_\boldmathsymbol{k}(\eta,\bar{\eta}) = \delta\left(\eta-\bar{\eta}\right),
\eeq
with the boundary conditions $\lim_{\eta\to \bar{\eta}}G^s_\boldmathsymbol{k}(\eta,\bar{\eta}) = 0$ and $ \lim_{\eta\to \bar{\eta}}G^{s,\prime}_\boldmathsymbol{k}(\eta,\bar{\eta})=1$.

\subsection{The scalar induced gravitational-wave signal}\label{sec:SIGW}
Considering now the effective energy density of the gravitational waves in the subhorizon region where one can use the flat spacetime approximation the latter can be recast as~\cite{Maggiore:1999vm} (see also Appendix of ~\cite{Isaacson:1968zza})
\bea
\label{rho_GW effective}
 \rhoGW (\eta,\boldmathsymbol{x}) & =  \frac{\Mp^2}{32 a^2}\, \overline{\left(\partial_\eta h_\mathrm{\alpha\beta}\partial_\eta h^\mathrm{\alpha\beta} +  \partial_{i} h_\mathrm{\alpha\beta}\partial^{i}h^\mathrm{\alpha\beta} \right)}\, .
\eea

In the radiation era, the scalar perturbations are in general exponentially suppressed due to diffusion damping~\cite{1980lssu.book.....P,1968ApJ...151..459S}, and therefore decouple in the late-time limit from the tensor perturbations. Therefore, considering only sub-horizon scales and neglecting the corresponding terms in \Eq{Tensor Eq. of Motion} (which now becomes a free-wave equation), the effective energy density of the gravitational waves reads as
\beq\label{rho_GW_effective_averaged}
\begin{split}
 \left\langle \rhoGW (\eta,\boldmathsymbol{x}) \right\rangle  & \simeq 2 \sum_{s=+,\times}\frac{\Mp^2}{32a^2}\overline{\left\langle\left(\nabla h^{s}_\mathrm{\alpha\beta}\right)^2\right \rangle }
 \\ & =   \frac{\Mp^2}{16a^2 \left(2\pi\right)^3} \sum_{s=+,\times} \int\mathrm{d}^3\boldmathsymbol{k}_1 \int\mathrm{d}^3\boldmathsymbol{k}_2\,  k_1 k_2  \\ & \times \overline{  \left\langle h^{s}_{\boldmathsymbol{k}_1}(\eta)h^{s,*}_{\boldmathsymbol{k}_2}(\eta)\right\rangle} e^{i(\boldmathsymbol{k}_1-\boldmathsymbol{k}_2)\cdot \boldmathsymbol{x}}\, .
 \end{split}
\eeq
where the bar denotes averaging over the sub-horizon oscillations of the tensor field and brackets mean an ensemble average. The factor of $2$ in the first line of \Eq{rho_GW_effective_averaged} stands for the fact that in the case of a free wave time derivatives are replaced with spatial ones.



In \Eq{rho_GW_effective_averaged} we see the presence of the equal time correlation function of the tensor perturbation field which actually defines the tensor power spectrum $\mathcal{P}_{h}(\eta,k)$ through the following expression:
\bea
\label{tesnor power spectrum definition}
\langle h^r_{\boldmathsymbol{k}_1}(\eta)h^{s,*}_{\boldmathsymbol{k}_2}(\eta)\rangle \equiv \delta^{(3)}(\boldmathsymbol{k}_1 - \boldmathsymbol{k}_2) \delta^{rs} \frac{2\pi^2}{k^3_1}\mathcal{P}^{(s)}_{h}(\eta,k_1),
\eea
where $s=(\times)$ or $(+)$. 

After a rather long but straightforward calculation and accounting for the fact that on the superhorizon regime $\Phi=2\mathcal{R}/3$~\cite{Mukhanov:1990me}, the tensor power spectrum $\mathcal{P}_{h}(\eta,k)$ can be recast as [see ~\cite{Kohri:2018awv,Espinosa:2018eve} for more details] 
\bea
\label{Tensor Power Spectrum}
\begin{split}
\mathcal{P}^{(s)}_h(\eta,k) = 4\int_{0}^{\infty} \mathrm{d}v\int_{|1-v|}^{1+v}\mathrm{d}u & \left[ \frac{4v^2 - (1+v^2-u^2)^2}{4uv}\right]^{2}\\ & \times I^2(u,v,x)\mathcal{P}_\mathcal{R}(kv)\mathcal{P}_\mathcal{R}(ku)\,,
\end{split}
\eea
with 
\bea
\label{I function}
I(u,v,x) = \frac{2}{3}\int_{x_\mathrm{d}}^{x} \mathrm{d}\bar{x}\, \frac{a(\bar{x})}{a(x)}\, k\, G_{k}(x,\bar{x}) F_k(u,v,\bar{x}).
\eea

Finally, defining the GW spectral abundance as the GW energy density per logarithmic comoving scale, combining \Eq{Tensor Power Spectrum} and \Eq{tesnor power spectrum definition} and plugging \Eq{tesnor power spectrum definition} into \Eq{rho_GW_effective_averaged} one obtains that 
\beq\label{Omega_GW}
\Omega_\mathrm{GW}(\eta,k)\equiv \frac{1}{\bar{\rho}_\mathrm{tot}}\frac{\mathrm{d}\rho_\mathrm{GW}(\eta,k)}{\mathrm{d}\ln k} = \frac{1}{24}\left(\frac{k}{\calH(\eta)}\right)^{2}\overline{\mathcal{P}^{(s)}_h(\eta,k)}.
\eeq

Then, one can show that at PBH formation time, namely at horizon crossing during the RD era, $\Omega_\mathrm{GW}$ can be recast as~\cite{Kohri:2018awv}
\beq\label{eq:Omega_GW_f}
\begin{split}
\Omega_\mathrm{GW}(\eta_\mathrm{f},k)  & = \frac{1}{12}\int_{0}^{\infty} \mathrm{d}v\int_{|1-v|}^{1+v}\mathrm{d}u \left[ \frac{4v^2 - (1+v^2-u^2)^2}{4uv}\right]^{2}\\ & \times \mathcal{P}_\mathcal{R}(kv)\mathcal{P}_\mathcal{R}(ku)  \left[\frac{3(u^2+v^2-3)}{4u^3v^3}\right]^{2} \\ & \times \biggl\{\biggl[-4uv + (u^2+v^2-3)\ln \left| \frac{3 - (u+v)^{2}}{3-(u-v)^{2}}\right|\biggr]^2  \\ & + \pi^2(u^2+v^2-3)^2\Theta(v+u-\sqrt{3})\biggr\}.
\end{split}
\eeq

At the end, accounting for entropy conservation between PBH formation time and today one can show that~\cite{Papanikolaou:2021aqq} 
\beq\label{Omega_GW_RD_0}
\Omega_\mathrm{GW}(\eta_0,k) = \Omega^{(0)}_r\frac{g_{*\mathrm{\rho},\mathrm{f}}}{g_{*\mathrm{\rho},0}}\left(\frac{g_{*\mathrm{S},\mathrm{0}}}{g_{*\mathrm{S},\mathrm{f}}}\right)^{4/3}\OmegaGW(\eta_\mathrm{f},k),
\eeq
where the subscript $0$ denotes the present value of the corresponding quantity, and $g_{*\mathrm{\rho}}$ and $g_{*\mathrm{S}}$ stand for the energy and entropy relativistic degrees of freedom. For our numerical numerical applications, we use $\Omega_\mathrm{rad,0}\simeq 10^{-4}$~\cite{Planck:2018vyg}, $g_{*\mathrm{\rho},0}\simeq g_{*\mathrm{S},0}= 3.36$, $g_{*\mathrm{\rho},\mathrm{f}}\simeq g_{*\mathrm{S},\mathrm{f}} = 106.75$~\cite{Kolb:1990vq}.

\begin{figure*}[ht]
\begin{center}
\includegraphics[width=0.795\textwidth]{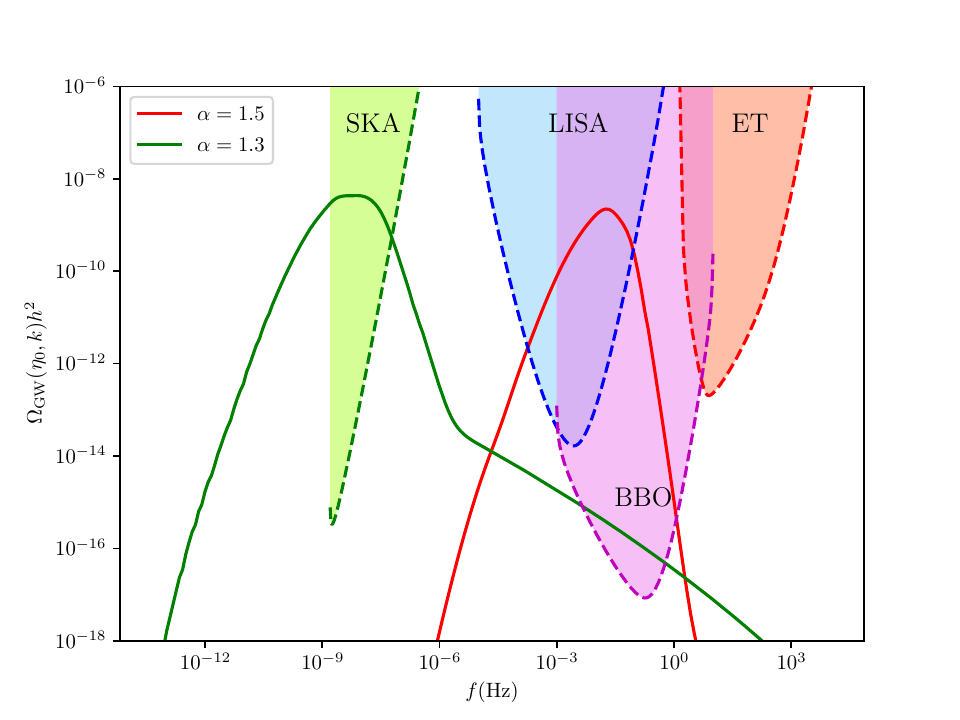}
\caption{{\it{The scalar induced gravitational-wave spectrum for different values of the non-canonicality parameter $\alpha$. With the green, blue, magenta and orange dashed curves we present respectively the SKA, LISA, BBO and ET gravitational-wave (GW) sensitivity curves. }}}
\label{fig:Omega_GW_vs_alpha}
\end{center}
\end{figure*}

In \Fig{fig:Omega_GW_vs_alpha}, we depict the GW spectra today for $\alpha=1.3$ and $\alpha = 1.5$ as a function of their frequency defined as $f=k/(2\pi a_0)$ with $a_0=1$ and for the model-parameters present in Table 1. We have superimposed as well the GW sensitivity curves of the Square Kilometer Array (SKA)~\cite{Janssen:2014dka}, Laser Space Inferometer Antenna (LISA)~\cite{Caprini:2015zlo,Karnesis:2022vdp}, Einstein Telscope (ET)~\cite{Maggiore:2019uih} and Big Bang Observer (BBO)~\cite{Harry:2006fi}. Interestingly, for $\alpha=1.5$ the peak frequency of the SIGW signal enters the LISA and BBO GW sensitivity bands whereas for $\alpha = 1.3$ the GW signal enters primarily within the SKA sensitivity with its tail slightly crossing the BBO curve. As expected, $\Omega_\mathrm{GW}$ follows the $k$ dependence of $\mathcal{P}^2_\mathcal{R}(k)$ as it can be speculated from \Eq{eq:Omega_GW_f}.

\section{Conclusions}\label{sec:conclusions}

PBHs constitute a general prediction of inflation. Traditionally, PBHs were studied within inflationary models with canonical kinetic terms. However, this needs not be the case, due to fine-tuning issues and predictions of large tensor modes which can be naturally addressed in theories of scalar fields with non-canonical kinetic terms. 

In this work, therefore, we study PBH formation within  ultra-slow-roll non-canonical inflation, working within a class of steep-deformed inflationary potentials which are compatible with natural values for the non-canonical exponents. Interestingly, by requiring that the USR phase should lead to an enhanced curvature power spectrum of the order of $\mathcal{O}(0.1)$ - in order to lead to PBH production - and accounting for the fact that inflation should not last more than $70$ e-folds we find that the non-canonical exponent $\alpha$ should be less than $1.5$ but larger than the canonical value $1$. These are quite natural values, and avoid the difficult to be justified regime $\alpha\gg1$ used in other non-canonical considerations.

Extending then our study to PBH formation within peak theory and accounting for the critical collapse scaling law for the PBH mass spectrum, we derive the contribution of PBHs to dark matter. We find a range of natural model parameters that can lead to the production of asteroid-mass PBHs, accounting for the totality of dark matter, as well as to formation of solar mass PBHs, which can be potentially detected by the LIGO/VIRGO experiments.

Moreover, we find that the enhanced cosmological perturbations which collapse to form PBHs can also lead to a stochastic gravitational-wave (GW) signal induced at second order in cosmological perturbation theory. Very interestingly, after extracting the respective GW spectra for different values of our model-parameters, we obtain GW signals within the GW sensitivity bands of SKA, LISA and BBO, thus potentially detectable by future GW experiments.

Finally, we should highlight as well some other portals through which one can test and constrain our non-canonical inflationary PBH formation scenario. In particular, one can extend our work by computing the bi/trispectrum  of the curvature perturbations studying possible non-Gaussian features within our inflationary set-up as well as their impact on the SIGW signal~\cite{Cai:2018dig}. Very interestingly, one should stress as well that PBHs within the lower asteroid mass window, namely with $m_\mathrm{PBH}\sim \mathcal{O}(10^{-17}M_\odot - 10^{-16}M_\odot)$, potentially produced within our model can be detected via Hawking radiation in near future gamma-ray telescopes~\cite{Ray:2021mxu}. This will be another test of our non-canonical inflationary scenario, besides the portal associated to gravitational waves.

\begin{acknowledgments}
T.P. acknowledges financial support from the Foundation for Education and European Culture in Greece. T.P. would like to thank as well the Laboratoire Astroparticule and Cosmologie, CNRS Université Paris Cité for giving him access to the computational cluster DANTE where part of the numerical computations of this paper were performed. The authors would like to acknowledge as well the contribution of the COST Action CA21136 ``Addressing observational tensions in cosmology with systematics and 
fundamental physics (CosmoVerse)''.  

\end{acknowledgments}

\bibliographystyle{JHEP} 
\bibliography{PBH}

\end{document}